\documentclass[preprint]{jpsj2}

\usepackage{graphicx}
\usepackage{amsmath}
\usepackage{amssymb}
\usepackage{bm}

\usepackage{color}

\title{%
First-Principles Computation of YVO$_3$;
Combining Path-Integral Renormalization Group
with Density-Functional Approach
} 
 
\author{%
Yuichi \textsc{Otsuka}$^{1,2}$
\thanks{Present address:
Synchrotron Radiation Research Center, 
Japan Atomic Energy Research Institute, 
Hyogo 679-5148, Japan }
\thanks{E-mail address: otsuka@spring8.or.jp}
and 
Masatoshi \textsc{Imada}$^{1,2,3}$
}

\inst{%
$^{1}$ 
Institute for Solid State Physics, 
University of Tokyo,
Kashiwanoha, Kashiwa, Chiba 277-8581, Japan\\
$^{2}$ 
PRESTO,
Japan Science and Technology Agency, 
4-1-8 Honcho, Kawaguchi, Saitama, Japan\\
$^{3}$
Depertment of Applied Physics,
University of Tokyo, 
7-3-1 Hongo Bunkyo-ku,
Tokyo 113-8656, Japan
}

\abst{%
We investigate the electronic structure of the transition-metal oxide
YVO$_3$ by a hybrid first-principles scheme.
The density-functional theory
with the local-density-approximation by using the local muffin-tin orbital basis
is applied to derive the whole band structure. The electron degrees of freedom
far from the Fermi level are eliminated by a downfolding procedure
leaving only the V $3d$ $t_{2g}$ Wannier band as the low-energy degrees of freedom, 
for which a low-energy effective model is constructed.
This low-energy effective Hamiltonian is solved exactly by
the path-integral renormalization group method.
It is shown that the ground state has
the $G$-type spin and the $C$-type orbital ordering in agreement with
experimental indications.
The indirect charge gap is estimated to be around 0.7 eV, which prominently 
improves the
previous estimates by other conventional methods.
}

\kword{
First-principles calculation, 
density-functional theory,
local-density-approximation,
path-integral renormalization group,
transition-metal oxide,
Mott transition,
orbital order,
charge gap
}

\begin{document}
\maketitle


\section{Introduction}

Recent intensive studies have revealed that
strongly correlated electron systems exhibit
a variety of distinguished physical properties.~\cite{MI-RMP}
Basic aspects of the correlation effects
have been studied mainly by simplified models, such as
the Hubbard, and Heisenberg models.
For example, at least some essence of the Mott transition, which is widely seen
in the strongly correlated electron systems,
is believed to be captured by the single-band Hubbard model.
However, if we focus on a variety of physical properties
in real materials, many challenging issues are found beyond the simplified models.
Especially in the systems where
spin and orbital are coupled, 
rich structure and various phenomena 
have attracted much interest.
When the strong Coulomb interaction suppresses charge fluctuations,
spin and orbital degrees of freedom play a crucial role.
In order to understand such realistic and complex systems,
computational methods from the first principles
offer promising approaches
because of their ability to treat 
charge, spin and orbital degrees of freedom
on equal footing with full fluctuation effects taken into account.

Computational methods for electronic structure calculations 
are roughly classified into two categories.
One is based on the density-functional theory (DFT)
supplemented, for example, 
with the local-density-approximation (LDA).~\cite{Hohenberg-Kohn,Kohn-Sham}
Since the LDA costs less computation time,
this scheme has proven its advantage in calculating
electronic structure of real materials.
However, the LDA essentially ignores 
spatial and dynamical fluctuations
which are relevant in the strongly correlated systems.
In addition, the LDA in general underestimates the charge gap.
The other approach is based on the wave function scheme.
The Hartree-Fock (HF) method often gives
a good starting point to discuss competition among ordered states,
while it tends to overestimate the charge gap.
The HF approximation itself is a rather crude one, where
spatial and temporal correlations are ignored.
However, one can improve the accuracy systematically 
within the wave function scheme
by expanding basis set, for instance, 
by introducing configuration interaction 
or 
by path-integral renormalization-group (PIRG) methods.~\cite{Kashima-JPSJ70}
The method along this line tends to need   
much longer computational time,
which prevents us from applying it directly to real materials.

Recently, a hybrid method called DFT-PIRG which combines the DFT with
a wave function method, namely PIRG has been proposed.~\cite{Imai-PRL95}
The DFT is utilized to first obtain the overall band structure extending to the
energy scale far from the Fermi level. Then the downfolding
procedure eliminates the high-energy degrees
of freedom far from the Fermi level by the renormalization process.
After this downfolding process,
an effective model for the low-lying
electronic structure is obtained, which is solved exactly 
by a low-energy solver.
The PIRG method is a promising choice
for a low-energy solver because it allows us
to treat spatial and dynamical fluctuations
in a controllable way.
It has indeed been applied to Sr$_2$VO$_4$~\cite{Imai-PRL95,Imai-JPSJ}
and has reproduced the basic experimental properties
including the fact that this compound is located 
just on the verge of the Mott transition 
with a tiny but nonzero gap amplitude ($\lesssim$ 0.1eV).
In the present work,
we study YVO$_3$
as another example of the charge-spin-orbital coupled system
by the DFT-PIRG method.

\begin{figure}[htbp]
 \begin{center}
 \includegraphics[width=14.5cm,clip]{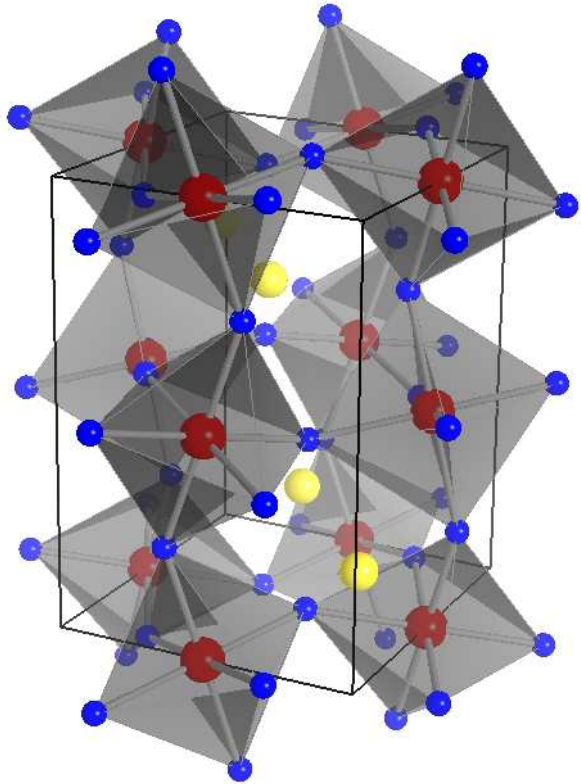}
  \caption{(Color online): 
  Illustration of the $Pbnm$ orthorhombic crystal structure of YVO$_3$. 
  Each V atom (large red sphere) is octahedrally surrounded  
  by O atoms (small blue spheres).
  Other (light yellow) spheres represents Y atoms.
  \label{fig:crystal} 
}
\end{center}
\end{figure}

YVO$_3$ belongs to the family of transition-metal oxides
with two valence electrons in the 3$d$ orbitals ($t_{2g}$ manifold).
The lattice structure is an orthorhombically distorted
perovskite with the space group $Pbnm$ 
(four vanadium sites in a unit cell)
at room temperatures.
The GdFeO$_3$-type distortion, rotation and tilting
of the VO$_6$ octahedra are present,
where the reduced V-O-V angle makes the narrow $t_{2g}$ bands.
With lowering the temperature,
it undergoes two successive phase transitions
in both spin and orbital sectors.
First, the $G$-type orbital ordering (OO)
appears at 200K with a structural change to the $P2_{1}/a$ symmetry,
where a site with the $d_{xy}$ and $d_{yz}$ orbitals occupied
and one with the the $d_{xy}$ and $d_{zx}$ are alternatively
arranged in three dimensions.
The magnetic structure  shows the $C$-type 
spin ordering (SO) below 116K,
where spins are aligned antiferromagnetically in the $a$-$b$ plane
and ferromagnetically along the $c$-axis.
With further lowering the temperature,
the SO and OO simultaneously change at 77K, and
the ground-state is the $C$-type OO with the $G$-type
SO.~\cite{Kawano,Miyasaka2003}
The crystal structure recovers the $Pbnm$ symmetry 
as illustrated in Fig.~\ref{fig:crystal}.
In the charge sector, 
YVO$_3$ is a typical Mott insulator with a large charge gap ($\sim$ 1eV).
This is partly attributed to a large GdFeO$_{3}$-type distortion, 
which reduces the band width effectively.
In addition, coupling to Jahn-Teller distortions is
important in determining the orbital states.

The relation between the SO and the OO
has been studied by the unrestricted
HF approximation.~\cite{Mizokawa1996,Mizokawa1999}
Although the HF result explains 
the spin and orbital orders in the ground-state
consistently with the experiments,
the magnitude of the charge gap is 
largely overestimated ($\sim$ 3.4eV)
compared to the experimental value ($\sim$ 1eV).
From the first-principles calculations,
the local spin density approximation (LSDA) and 
the generalized gradient approximation (GGA) 
have been applied with the full-potential
linearized augmented-plane-wave method.~\cite{Sawada1996,Sawada1998}
The LSDA results have shown the metallic ground state.
On the other hand, the ground state is insulating with GGA,
but it failed to reproduce the correct SO and OO.
In addition, the gap amplitude is underestimated (0.009eV).
The LDA+$U$ method has succeeded in describing 
the correct ground state~\cite{Sawada1998,Fang2003}. 
However, aside from the issue of avoiding the double counting
of the Coulomb interaction generally known in LDA+$U$ calculation, 
this ground state is obtained
by adjusting the parameter $U$ 
to reproduce the experimental band gap. 
It is desired to estimate the screened Coulomb interaction itself from 
the first principle. In addition, it is also important to
go beyond the single-Slater-determinant approximation 
to examine effects of quantum fluctuations accurately,
and to estimate spatial and temporal fluctuations from the first principles. 
This is because orbital-spin fluctuations may seriously alter the ground state 
through strong correlations combined with quantum
fluctuations. 

The purpose of the present work is
to study the ground state of YVO$_3$ from the first-principles
calculation at a quantitative level.
We will discuss how the SO and OO are reproduced
with the DFT-PIRG scheme and estimate the charge gap.
The basic procedures are the same as the case of Sr$_2$VO$_4$.
Readers are referred to refs.~\citen{Imai-PRL95,Imai-JPSJ} 
for more detailed description of this scheme.
Here we note that the target materials are quite different,
while the procedures are similar.
In contrast to Sr$_2$VO$_4$,
YVO$_3$ is a cubic perovskite with the GdFeO$_{3}$-type distortion,
where its isotropic three dimensionality or mixing between $t_{2g}$ orbitals
may show different aspects.
The inter-orbital Coulomb interaction and the Hund's rule coupling 
are considered to be more effective for the $3d^{2}$ configuration
of YVO$_3$.
Furthermore, YVO$_3$ has much larger Mott gap ($\sim$ 1eV).
To examine the applicability of the DFT-PIRG method,
it is desired to test whether it works 
in such completely different systems.

In Sec. II, we derive the effective low-energy model.
In Sec.III, the Hartree-Fock solution to this model is considered.
We discuss results of the PIRG calculation in Sec. IV. 
Section V is devoted to summary.

\section{Constructing Effective Model}

In many cases of the strongly correlated electron systems,
the bands close to the Fermi level are
relatively narrow and well separated from
the bands far from the Fermi level.
In fact, for the transition metal oxides,
the $3d$ bands are mostly responsible
for the low-energy excitations.
Compared to the width of the $3d$ bands,
other bands are located far from the Fermi level.
Furthermore, if the crystal-field splitting is strong,
it may be sufficient to consider 
only either $e_{g}$ or $t_{2g}$ bands for the low-energy degrees of freedom.
This hierarchy structure in energy enables us to
treat the system in a hybrid scheme.

The electron correlation effects and dynamical fluctuations 
are important near the Fermi level, whereas
the electronic structure far from the Fermi level 
is expected to be well reproduced by LDA. 
Therefore, in the first step of our hybrid scheme, 
the whole electronic structure is calculated 
from the density functional approach by
employing LDA based on the linear muffin-tin orbital (LMTO)
method.~\cite{OKAnderson,Gunnarsson} 
The LDA enables us to calculate 
the whole electronic structure 
within the presently available computer power.  
Then we eliminate the electronic degrees freedom 
far from the Fermi level.
This is done by the renormalization (downfolding) procedure,
where the electron correlation effects are properly considered 
in the calculation of the screening and self-energy effects. 
After the downfolding procedure, the low-energy degrees of freedom 
isolated near the Fermi level are extracted 
in the effective Hamiltonian, 
which will be solved accurately 
in the PIRG method.

\begin{figure}[htbp]
 \begin{center}
  \includegraphics[width=8.5cm,clip]{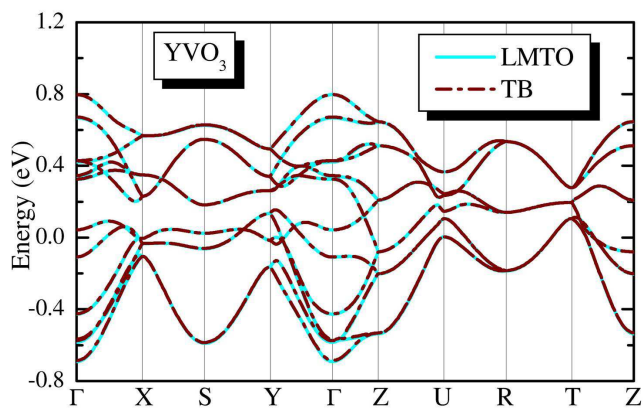}
  \caption{(Color online):
  Comparison of the band structure of $3d$ $t_{2g}$ orbitals
  computed from LMTO calculations (solid light blue lines)
  with the 
  downfolded tight-binding model (dashed-dotted brown lines)
for YVO$_3$.
  \label{fig:bands} 
  }
 \end{center}
\end{figure}

Let us explain the whole procedure developed in the literature~\cite{Imai-PRL95,Imai-JPSJ} 
by extending it specifically for the present compound YVO$_3$.
At first,
we compute the LDA band structure from the LMTO basis.
Since YVO$_3$ has sufficiently large 
crystal-field splitting ($\sim 0.1$eV)
due to a large Jahn-Teller distortion,
we focus on three low-energy bands
which mainly consist of the $3d$ $t_{2g}$ orbitals.
The whole Hilbert space is now spanned by the LDA basis functions for the 
LMTO $t_{2g}$ bands $\{|d\rangle\}$ and the set of the rest basis $\{|r\rangle\}$. 
Then we eliminate the subspace $\{|r\rangle\}$ and reduce the Hilbert space
to the subspace which only consists of the set $\{|d\rangle\}$.~\cite{Igor2004,Imai-PRL95}
The Hamiltonian matrix is now spanned only within this restricted Hilbert subspace,
which provides us with the low-energy tight-binding effective Hamiltonian defined on the
Wannier basis for the $3d$ $t_{2g}$ bands.
Note that
the band structure of the effective tight-binding Hamiltonian 
defined only on the low-energy subspace well reproduces 
the $t_{2g}$ part of LDA bands computed from the whole LMTO basis
as is illustrated in Fig.~\ref{fig:bands}.

Next, the parameter values for the final effective low-energy Hamiltonian 
for the $3d$ $t_{2g}$ Wannier bands become
renormalized by the interaction between electrons in the subspace $\{|r\rangle\}$ 
and 
those of $3d$ $t_{2g}$ near the Fermi level in the subspace $\{|d\rangle\}$. 
The renormalization appears both in the screening of the interaction parameters
and the renormalization of the kinetic energy part. 
These two procedures are briefly outlined in the following paragraphs.


Here we sketch out the derivation of 
the effective Coulomb interaction among the $3d$ $t_{2g}$ electrons,
which consists of two steps.
In the first step, by using the standard constrained-LDA (c-LDA) method,
we compute the interaction matrix $W_{r1}$ without
effects of hybridization between $3d$ and non-$3d$ orbitals.
This part takes into account only the screening caused by
the redistribution of the non-$3d$ electrons.
From this c-LDA procedure, we obtain the screening of the Coulomb interaction 
between two electrons in the $t_{2g}$ Wannier 
band caused by the polarization of electrons at the non-3$d$ orbitals in the LMTO basis.  
In the second step, we consider the screening of the Coulomb interaction between $t_{2g}$ electrons 
caused by polarization of electrons on the 3$d$ LMTO basis other than the $t_{2g}$ basis  
(namely caused by the $e_g$ component and the 3$d$ LMTO atomic orbital component residing
in the oxygen 2$p$ Wannier band)
in the random-phase-approximation (RPA) scheme.
By taking $W_{r1}$ obtained from the c-LDA scheme as a starting Coulomb interaction,
we can write the RPA screening as
\begin{equation}
 W_{r}(\omega) = \frac{W_{r1}}{1 - P_{dr}(\omega)W_{r1}},
\end{equation}
where $P_{dr}$ denotes polarization of electrons in the $3d$ LMTO basis
but in the non-$t_{2g}$ Wannier bands.
The total polarization of the $3d$ bands is given by
\begin{equation}
 P_{d}(\omega) = P_{dr}(\omega) + P_{t2g}(\omega),
\end{equation}
where $P_{t2g}$ represents contribution from $3d$ $t_{2g}$ Wannier bands.
Then, we notice that the following identity holds,
\begin{equation}
 W(\omega) = \frac{W_{r1}}{1-P_{d}(\omega) W_{r1}}
           = \frac{W_{r}(\omega) }{1-P_{t2g}(\omega)W_{r }(\omega)}.
\end{equation}
It implies that $W_{r}(\omega)$ can be considered to be
the effective Coulomb interaction among the $3d$ $t_{2g}$ electrons.
Although this partially screened interaction $W_{r}(\omega)$
in general is dynamical,
its frequency dependence is small within the 
energy scale of the $t_{2g}$ bandwidth.
Thus we can safely take the zero-frequency limit
$W_{r}(\omega=0)$ as the effective Coulomb interaction
to construct the low-energy Hamiltonian,
which will be treated exactly by the PIRG method.
We note that 
this separability of the screening effects
enables the downfolding procedure.
The reason why we perform two steps in calculating the screening is that
the c-LDA is less computer-time consuming, whereas it does not take into account the 
frequency dependence.  On the other hand, RPA is more time consuming, whereas
it is to some extent able to consider dynamical fluctuations, 
which is important near the  Fermi level. 

In the downfolding process,
the kinetic-energy part is also modified
through the self-energy $\Sigma(k,\omega)$.
Such a self-energy is mostly momentum independent
and its imaginary part is negligible 
at low energy.
Thus the self-energy effect mainly contributes to
the renormalization factor $Z$ given by,
\begin{equation}
 Z = \left[ 
      1 - \partial \textrm{Re} \Sigma / \partial \omega |_{\omega=0} 
     \right]^{-1},
\end{equation}
which is obtained by the numerical estimates from 
the $\omega$-dependence of the self-energy $\Sigma(\omega)$.
This means that the self-energy effect reduces the band width,
leading to the kinetic-energy part renormalized by $Z$.
The self-energy $\Sigma(k,\omega)$ can be evaluated
in the GW approximation, and
the renormalization factor $Z$ has been calculated for
several transition metal oxides.
We take $Z=0.8$, which is a typical value among these materials.
For more complete description of the downfolding procedure, 
readers are referred to 
refs.~\citen{Imai-PRL95},~\citen{Imai-JPSJ},~\citen{Igor2005-prl},~\citen{Igor2005-prb} and~\citen{Aryasetiawan-prb2005}.

\begin{figure}[htbp]
 \begin{center}
  \includegraphics[width=8.5cm,clip]{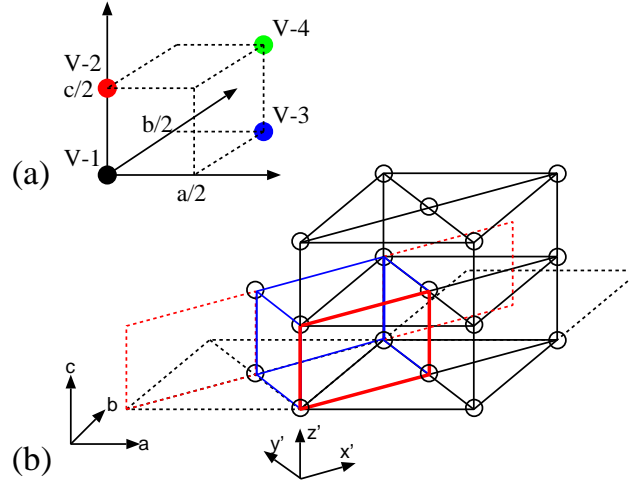}
  \caption{(Color online):
(a) The definition of the unit cell.
    Each unit cell has 4 vanadium atoms(V-1, V-2, V-3 and V-4).
(b) The lattice structure of YVO$_3$. 
    The thick (red) solid line indicates a unit cell.
  \label{fig:lattice-01} 
  }
 \end{center}
\end{figure}

After eliminating the high-energy part,
the effective Hamiltonian is reduced to
a three-band Hubbard model
in three dimensions:
\begin{equation}
 \mathcal{H}
=
 \mathcal{H}_{\text{k}}
+ \lambda \mathcal{H}_{\text{U}},
\end{equation}
where 
$\mathcal{H}_{\text{k}}$ and 
$\mathcal{H}_{\text{U}}$ 
denote 
the kinetic and interaction energy terms,
respectively and $\lambda$ is a tuning
parameter which controls the overall strength of the interaction.
The realistic value corresponds to $\lambda=1$.
Due to the lattice distortion,
each unit cell has four 
vanadium sites.
The definition of the unit cell is shown in
Fig.\ref{fig:lattice-01}.
The kinetic part of the Hamiltonian is given by
a tight-binding model,
\begin{align}
 \mathcal{H}_{\text{k}}  
 &=
   \sum_{\sigma}
   \sum_{i, j} 
   \sum_{l, l^{\prime}} 
   \sum_{m, m^{\prime}} 
 c_{i l m \sigma}^{\dagger} 
 t_{i j l l^{\prime} m m^{\prime}}
 c_{j l^{\prime} m^{\prime} \sigma},
\end{align}
where $c_{i l m \sigma}^{\dagger}(c_{i l m \sigma})$ denotes 
the creation (annihilation) operator
for spin $\sigma = \uparrow,\downarrow$ at $m$-th orbital
of site $l$ in the $i$-th unit cell 
and 
$t_{i j l l^{\prime} m m^{\prime}}$ is
the overlap integral.
We note that
the orbitals which we consider here
are ``local $t_{2g}$ orbitals'',
which is constructed basically 
from the local crystal-field
of the VO$_6$ octahedral frame.
Because of the tilting distortion of the VO$_6$ octahedra,
each orthogonal orbital is a mixture of
$d_{xy}$, $d_{yz}$, $d_{z^{2}}$, $d_{zx}$ and $d_{x^{2}-y^{2}}$
in the original coordinate.
However, the $t_{2g}$ components are still dominant
due to the large crystal-field splitting.

We choose a base $|l,m \rangle$ that diagonalizes the on-site
matrix elements $t_{i i l l m m^{\prime}}$~\cite{Yamasaki}.
The expansion coefficients of the unitary transformations over the
real harmonics 
($|xy \rangle$, $|yz \rangle$, $|z^{2} \rangle$, $|zx \rangle$, $|x^{2}-y^{2} \rangle$)
in the orthorhombic coordinate
frame are given as follows:
\begin{equation}
 \begin{pmatrix}
  |l,1 \rangle \\
  |l,2 \rangle \\
  |l,3 \rangle \\
 \end{pmatrix}
=
\mathcal{M}_{l}
\begin{pmatrix}
 |xy \rangle \\
 |yz \rangle \\
 |z^{2} \rangle \\
 |zx \rangle \\
 |x^{2}-y^{2} \rangle
\end{pmatrix},
\end{equation}
\begin{gather}
 \mathcal{M}_{1}=
\begin{pmatrix}
  -0.2302 &  -0.3619 &  -0.3785 &   0.8195 &   0.0355 \\
  -0.8913 &   0.2435 &   0.0527 &  -0.1027 &  -0.3646 \\
   0.0605 &   0.7922 &   0.2544 &   0.4720 &   0.2851 \\
\end{pmatrix},\\
 \mathcal{M}_{2}=
\begin{pmatrix}
  -0.2304 &   0.3617 &  -0.3784 &  -0.8195 &   0.0355 \\
  -0.8912 &  -0.2434 &   0.0527 &   0.1030 &  -0.3647 \\
   0.0603 &  -0.7924 &   0.2544 &  -0.4718 &   0.2850 \\
\end{pmatrix},\\
 \mathcal{M}_{3}=
\begin{pmatrix}
  -0.2318 &   0.8193 &  -0.3784 &  -0.3614 &  -0.0349 \\
  -0.8909 &  -0.1044 &   0.0533 &   0.2439 &   0.3648 \\
   0.0601 &   0.4719 &   0.2544 &   0.7924 &  -0.2849 \\
\end{pmatrix},\\
 \mathcal{M}_{4}=
\begin{pmatrix}
  -0.2302 &  -0.8198 &  -0.3783 &   0.3614 &  -0.0357 \\
  -0.8913 &   0.1026 &   0.0527 &  -0.2437 &   0.3646 \\
   0.0608 &  -0.4715 &   0.2546 &  -0.7924 &  -0.2852 \\
\end{pmatrix},
\end{gather}
where we can see that
the orbitals $m=1$ and $m=3$ mainly correspond to $d_{zx}$  and $d_{yz}$
($d_{yz}$ and $d_{zx}$) orbitals, respectively for $l=1,2 (3,4)$ and
the orbitals $m=2$ mainly consist of $d_{xy}$.
The diagonalized matrix has $(m,m)$ elements given by
\begin{gather}
 t_{i i l l m m^{\prime}} =
 \begin{pmatrix}
  -0.2711 & -0.0000 &  0.0000\\ 
   0.0000 & -0.2175 &  0.0000\\
   0.0000 & -0.0000 & -0.1054
 \end{pmatrix}, \label{eq:cfs}
\end{gather}
within each vanadium atom
(namely only diagonal elements for $i$ and $l$ are shown here).
This matrix elements do not depend on $i$ and $l$.
We summarize other off-diagonal matrix elements in
Appendix~\ref{sec:transfer}.
The lattice distortion brings about level splittings
among $t_{2g}$ orbitals, which is estimated to be 0.17 eV
as seen in eq.(\ref{eq:cfs}).
We take into account up to the third-nearest-neighbor transfers because 
the transfers beyond them are negligible.


The Coulomb interaction term is considered 
in each vanadium site:
\begin{align}
 \mathcal{H}_{U} 
&=
\sum_{i,l, m, \sigma}
U_{m m} 
n_{i l m \sigma} n_{i l m -\sigma} 
+ 
\sum_{i,l, m \ne m^{\prime}, \sigma}
U_{m m^{\prime}} 
n_{i l m \sigma} n_{i l m^{\prime} -\sigma} 
+
\sum_{i,l, m \ne m^{\prime}, \sigma}
\left( U_{m m^{\prime}}  - J_{m m^{\prime}}\right)
n_{i l m \sigma} n_{i l m^{\prime} \sigma} \nonumber \\
-&
\sum_{i,l, m \ne m^{\prime}, \sigma}
J_{m m^{\prime}}
\left(
c_{i l m           \sigma}^{\dagger}
c_{i l m          -\sigma}
c_{i l m^{\prime} -\sigma}^{\dagger}
c_{i l m^{\prime}  \sigma}
-
c_{i l m           \sigma}^{\dagger}
c_{i l m          -\sigma}^{\dagger}
c_{i l m^{\prime} -\sigma}
c_{i l m^{\prime}  \sigma}
\right),
\end{align}
where
$U_{m m}$ and
$U_{m m^{\prime}}$
denote intra- and inter-orbital Coulomb repulsion,
and $J_{m m^{\prime}}$ is the Hund's rule coupling.
We have checked that the long-range interaction is 
irreverent for this system.
The value of the interaction is shown in Table~\ref{Table1}.

\begin{table}
\begin{center}
\begin{tabular}{|c|c c c|c c c|c c c| }
\hline
& $U_{11}$ & $U_{22}$ & $U_{33}$ & $U_{12}$ & $U_{13}$ & $U_{23}$ & $J_{12}$ & $J_{13}$ & $J_{23}$ \\
\hline
energy(eV) & 3.178  & 3.235  & 3.232  & 1.961  & 1.955  & 1.985  & 0.633  & 0.627  & 0.633  \\
\hline
\end{tabular}
\end{center}
\caption{
The interaction energy obtained by the downfolding procedure.
}
\label{Table1}
\end{table}

\section{Hartree-Fock Calculation}
\begin{figure}[htbp]
 \begin{center}
  \includegraphics[width=8.5cm,clip]{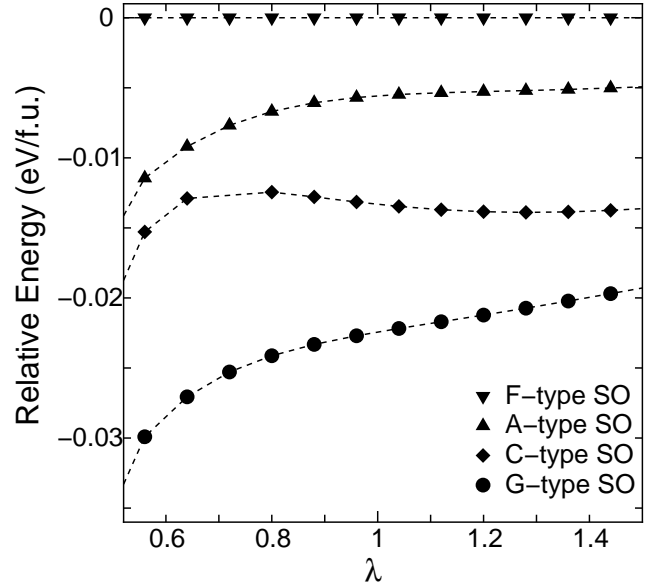}
\end{center}
  \caption{
 Energy difference between the FM and various AF solutions
 as functions of $\lambda$ in the HF solution.
 The closed circles, diamonds and upper triangles
 represent the results of $G-$, $C-$ and $A-$type AF solutions,
 respectively. The lower triangles is the results of the FM.
  \label{fig:HF-01} 
  }
\end{figure}

\begin{figure}[htbp]
 \begin{center}
 \includegraphics[width=7.5cm,clip]{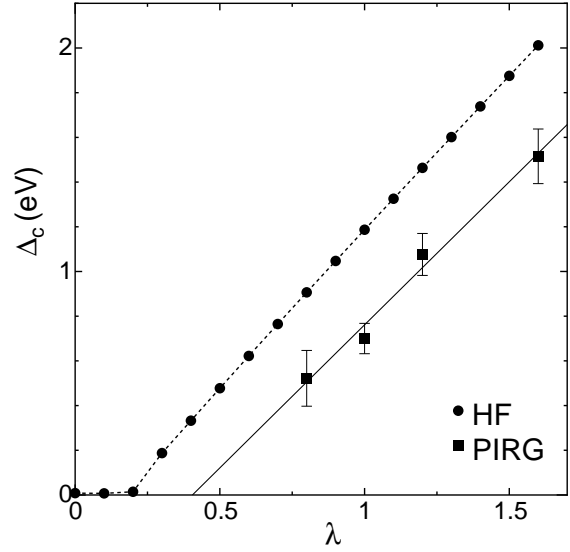}
  \caption{
  The $\lambda$ dependence of the charge gap $\Delta_{c}$
  for the HF ($L=1$) solutions (circles) and
  the PIRG results (squares).
  The solid line is the  least-square fit
  to the PIRG results.
  \label{fig:gap} 
}
\end{center}
\end{figure}

As a starting point of the PIRG calculation,
we apply the Hartree-Fock approximation.
The effective model is decoupled
to a mean-field Hamiltonian
in the usual manner.
Concentrating on homogeneous solutions,
we take the order parameter as
\begin{align}
 b_{l m m^{\prime} \sigma}
\equiv&
\frac{1}{N}
\sum_{i}
 \langle c_{i l m \sigma}^{\dagger} c_{i l m^{\prime} \sigma} \rangle,
\end{align}
where $N$ denotes number of unit cells.
Assuming a specific SO as an initial state,
we calculate the HF self-consistent equations
until the energy converges.
During the HF calculations, 
each SO is retained
when the interactions are large,
while the spin structure relaxes into a paramagnetic solution
for small interactions.
In Fig.~\ref{fig:HF-01},
we show the relative HF energy for each SO
measured from the ferromagnetic (FM) solution.
It is consistent with the experimental result 
and the previous HF calculations~\cite{Mizokawa1996}
that the $G$-type SO is the lowest in energy.
The energy of the $G$-type SO solution is close to 
that of the $C$-type SO,
which naturally explains the fact that
YVO$_3$ is located near the phase boundary
between the  $G$-type and $C$-type SO.~\cite{Miyasaka2002}
The HF ground state with $G$-type SO solution 
also gives
the $C$-type OO.
With these SO and OO,
we also calculate the HF charge gap 
$\Delta_{\text{c}}^{\text{HF}}$ (Fig.\ref{fig:gap}),
which gives 1.19eV for the realistic value ($\lambda =1.0$).
Despite the HF method in general often overestimates the charge gap,
our result is very close to the experimental one ($\sim 1$ eV).
Indeed, the gap amplitude is overestimated by 2eV
in ref.~\citen{Mizokawa1996}, although it
also employs the Hartree-Fock approximation as well.
The difference comes from the difference in the effective low-energy model.
It is crucially important to derive a reliable low-energy model 
from the first principles.
Our effective model obtained by the downfolding procedure
has an advantage in reproducing both the order patterns
and the gap amplitude even at the HF level.
We discuss the charge gap in detail below.

\section{PIRG Calculation}
Here we briefly summarize the procedure of the PIRG calculation.~\cite{Kashima-JPSJ70}
The ground state wave function is extracted from
a proper initial wave function $| \Phi \rangle$
by following the principle that
$
\lim _{p \rightarrow \infty} 
\left[
\exp \left(-\tau \mathcal{H} \right)
\right]^{p} | \Phi \rangle
$ generates the ground state where $\tau$ represents a small number
such that we can use the Suzuki-Trotter decomposition.
The interaction terms are decoupled 
by the Hubbard-Stratonovich (HS) transformation.
The path integral is expressed by the summation over the HS variables.
The variational wave function is composed of
$L$ nonorthogonal Slater determinant basis functions as
$| \Phi \rangle _{L} = \sum_{i=1}^{L} c_{i} | \phi _{i}\rangle$.
For a fixed $L$, 
the  coefficients $c_{i}$ and the choice of basis $| \phi _{i}\rangle$
are optimized.
Note that the case of $L=1$ corresponds to the HF solution.
The spatial and dynamical fluctuations 
are taken into account
by increasing a number of Slater-determinant basis functions $L$.
For sufficient large $L$, the ground-state energy decreases
linearly as a function of the energy variance.
This linear behavior ensures valid extrapolation
to the full Hilbert space.

\begin{figure}[htbp]
 \begin{center}
(a)
 \includegraphics[width=7.5cm,clip]{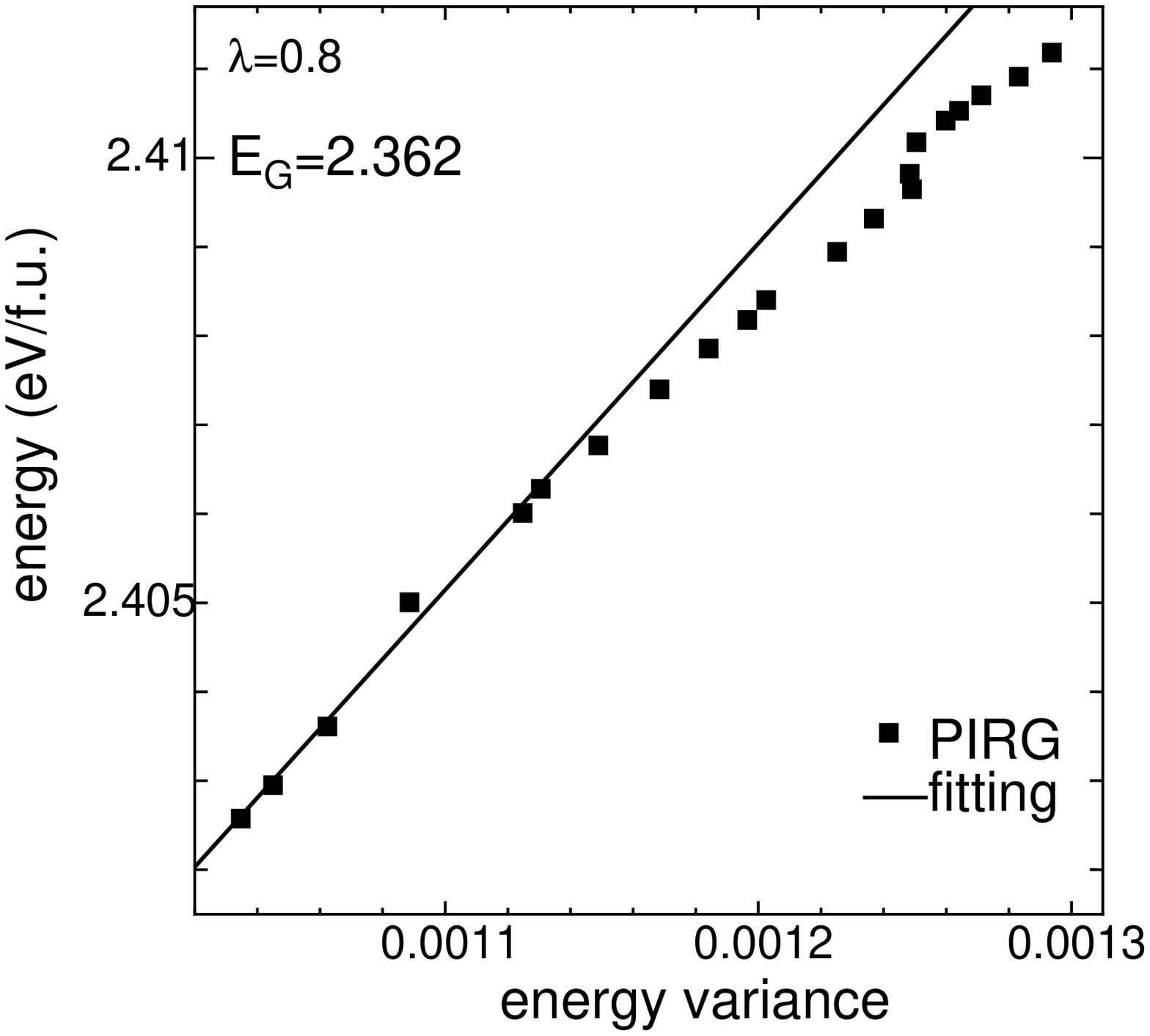}

(b)
 \includegraphics[width=7.5cm,clip]{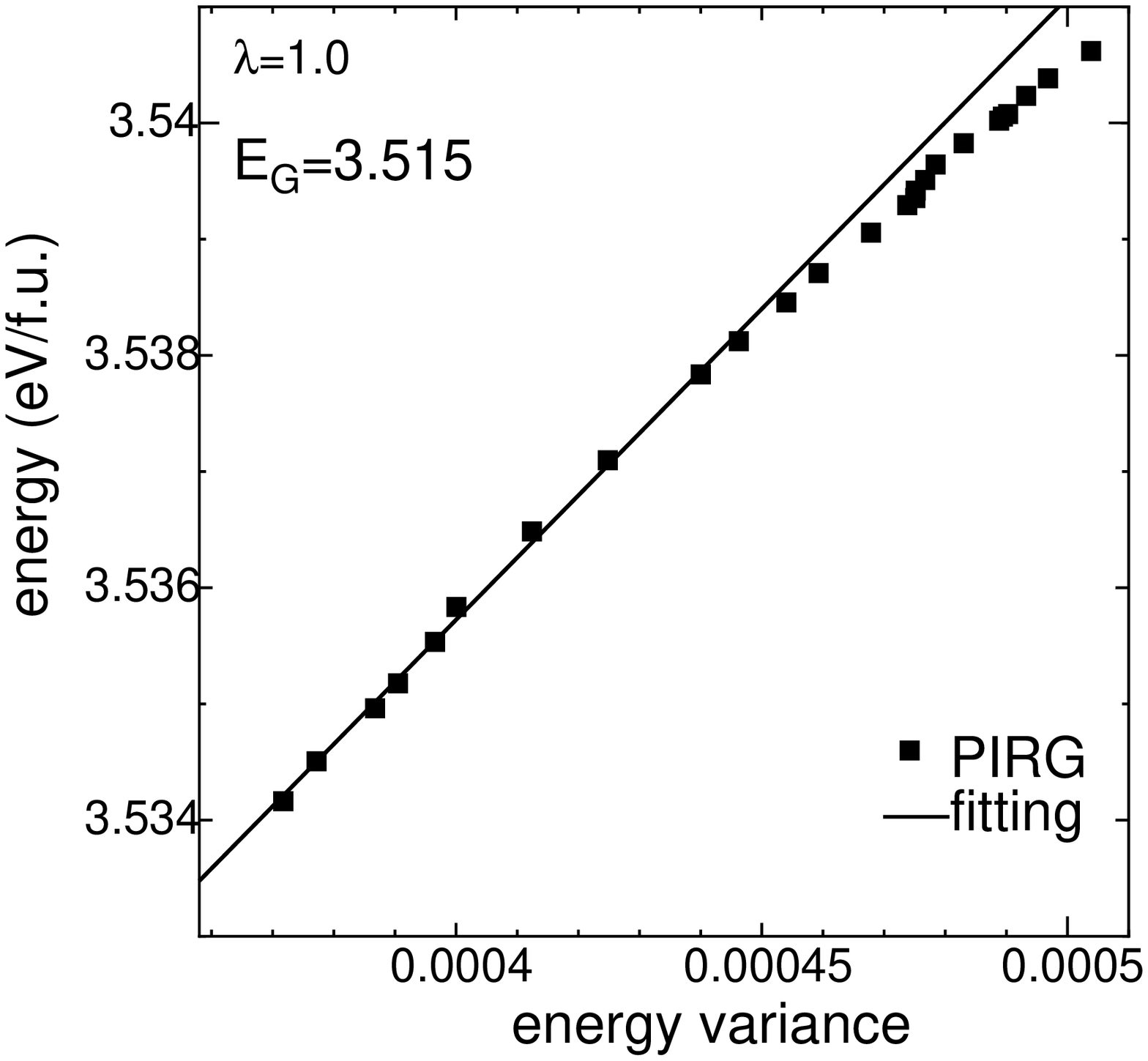}

(c)
 \includegraphics[width=7.5cm,clip]{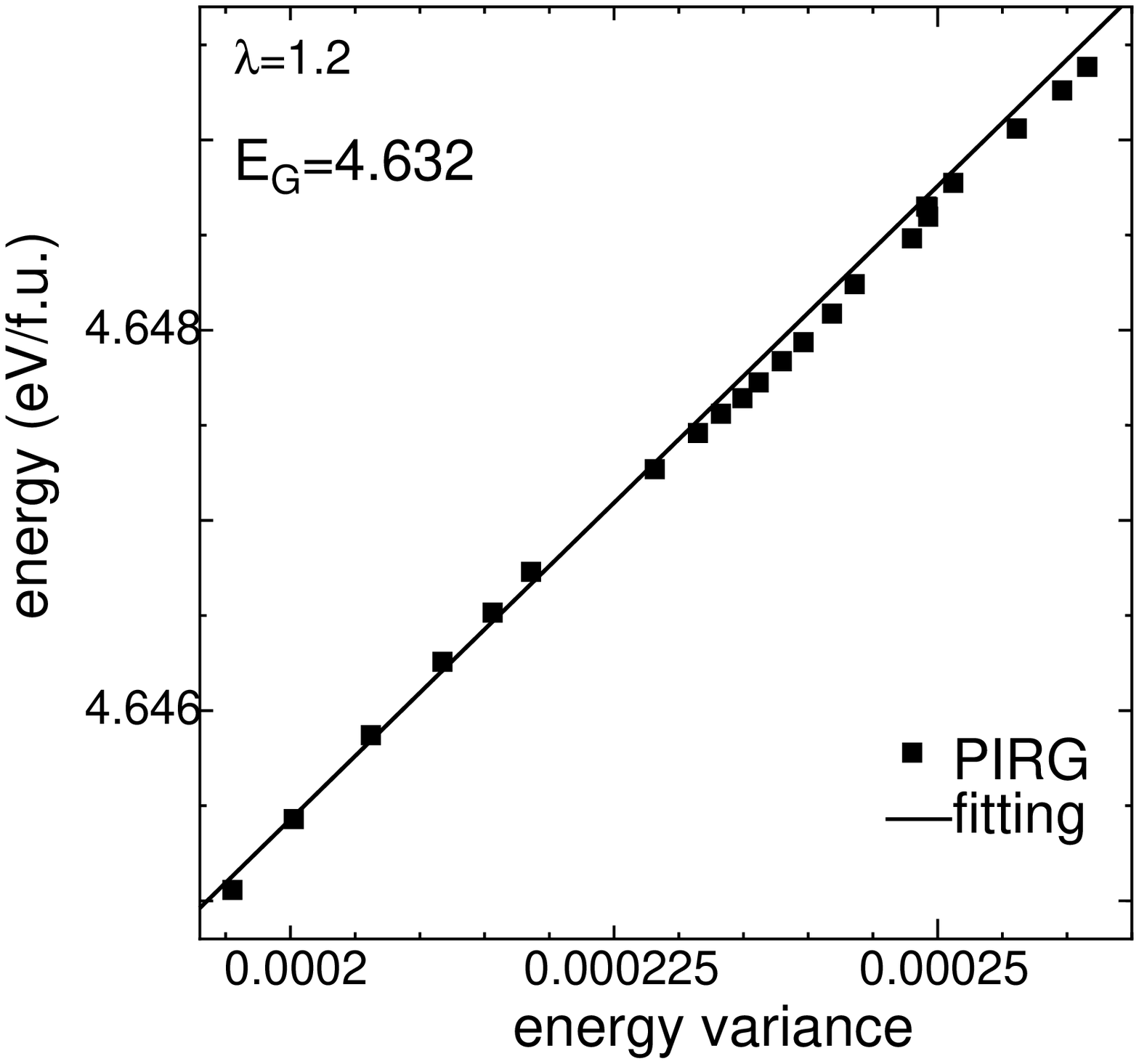}

  \caption{
  The ground-state energy per unit as function of energy variance 
  calculated by PIRG
  for (a) $\lambda$=0.8, (b) $\lambda$=1.0 and (c) $\lambda$=1.2.
  $E_{\text{G}}$ denotes the extrapolated value of energy
  to the energy variance.
  \label{fig:energy-PIRG} 
}
\end{center}
\end{figure}

\begin{figure}[htbp]
 \begin{center}
 \includegraphics[width=7.5cm,clip]{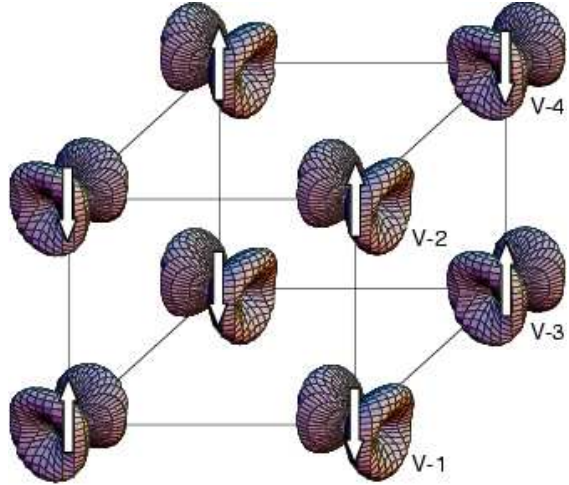}
  \caption{(Color online):
  Ordered spin- and orbital- patterns 
  in the ground state of the PIRG solution.
  The arrows represent the magnetic local moment at each vanadium
  atom. 
  The orbital states are shown in the form of 
  the spatial electron distribution.
  \label{fig:HF-02} 
}
\end{center}
\end{figure}

For YVO$_3$, 
the wave function of the $G$-type SO solution
obtained by HF calculations
as an initial state of the PIRG calculation gives the lowest energy state.
We have performed calculations up to $L=192$ on 
$N = 2 \times 2 \times 2$ unit cells (32 vanadium atoms).
We discuss system-size dependence later.
In Fig.\ref{fig:energy-PIRG},
we can see an example that this linear extrapolation
works well in a controllable way.
We also check that 
the $G$-type SO and $C$-type OO are preserved
during the extrapolation process, while their
amplitudes are slightly modified. 
The spin and orbital configurations 
in the ground state are 
illustrated in Fig.~\ref{fig:HF-02}

We calculate the total energy $E(M_{\uparrow}, M_{\downarrow})$
for the system with $M = M_{\uparrow}+M_{\downarrow}$ electrons, 
where $M_{\sigma}$ denotes number of electrons with spin $\sigma$.
Then we obtain the chemical potential as follows:
\begin{align}
 E_{+} =& E(4N+1, 4N+1) \\
 E_{0} =& E(4N, 4N) \\
 E_{-} =& E(4N-1, 4N-1) \\
\mu_{+} =& \frac{\Delta E}{\Delta M} = \frac{E_{+}-E_{0}}{2}\\
\mu_{-} =& \frac{\Delta E}{\Delta M} = \frac{E_{0}-E_{-}}{2},
\end{align}
where $E_{0}$ is the energy of the $1/3$-filled system
which corresponds to YVO$_3$.
The charge gap $\Delta_{\text{c}}$ is estimated as 
a difference between these two chemical potentials,
\begin{equation}
 \Delta_{\text{c}} = \mu_{+} - \mu_{-}.
\end{equation}
We note that the gap obtained by the above procedure
is the indirect one
that should give the lower bound of the direct gap.
Figure~\ref{fig:gap} shows $\lambda$ dependence of the charge gap
calculated by HF and PIRG.
As the result of fluctuation,
$\Delta_{\text{c}}$ of PIRG is reduced by 30-40\% 
in comparison with the HF results.
We also see that $\Delta_{\text{c}}$ behaves linearly
as a function of $\lambda$, which is similar to the HF results.
For the realistic value ($\lambda =1.0$),
the charge gap is 0.70 $\pm$ 0.07 eV, 
which seems to be slightly smaller than
the experimental optical gap.~\cite{Miyasaka2002}
This is partly because that
the PIRG calculation gives the indirect gap in general,
while the optical gap is the direct one.
The difference between 
the indirect and direct gap is estimated as
$\sim$ 0.1eV in the HF solution.
Since the PIRG calculation is performed on a finite-size system,
the charge gap may further be reduced with a finite-size scaling.
Because of the limitation of computer resources, for the moment,
results on finite-size scaling are not available.
However, 
we have checked that the charge gap is insensitive
to the system size in the HF calculations.
It may suggest that 
the charge gap is determined 
from a local origin.
We believe that 
the present result is close to the thermodynamic limit.
Our result is quantitatively consistent with the experimental results
as compared to those of HF(3.4eV)~\cite{Mizokawa1996} 
or GGA (0.009eV).~\cite{Sawada1996}
We stress that the present results are obtained from
the first-principles calculation 
without any adjustable parameters.

\section{Summary}
In summary, 
the transition metal oxide YVO$_{3}$
is investigated by the DFT+PIRG scheme.
First we use the DFT method to construct the effective low-energy
Hamiltonian which consists of the tight-binding part and
the effective screened Coulomb interactions.
The PIRG method is employed as a low-energy solver
that can fully take into account spatial and dynamical fluctuations.
The obtained spin and orbital state is consistent with the experiment.
The indirect-charge gap is estimated to be 0.70 $\pm$ 0.07 eV, 
which is smaller than the inferred experimental optical (direct) gap,
but is consistent each other in terms of the difference between direct and indirect gaps.  It has prominently improved the estimation compared to 
the LDA or GGA method.
Our results are all consistent with the available experimental results.
It is desired to measure the indirect gap 
to compare with the present result.
The present application 
to YVO$_3$
further proves that
DFT+PIRG offers a quantitatively precise first-principles scheme
for strongly correlated electron systems.
The method establishes it as a standard scheme 
for reliable estimates of the electronic structure
when the lattice structure is given.

\begin{acknowledgments}
The authors would like to thank 
Igor Solovyev for providing us 
with the data for the downfolded Hamiltonian 
and valuable suggestions.
We are also grateful to Y. Imai for fruitful discussions.
YO acknowledges helpful comments with A. Yamasaki.
This work is partially supported by grant-in-aids for
scientific research from 
Ministry of Education, Culture, Sports, Science and Technology
under the grant numbers 16340100 and 17064004.
A part of our computation has been done
at the supercomputer center of 
the Institute for Solid State Physics,
University of Tokyo.
\end{acknowledgments}

\appendix
\section{Transfer Integrals}
\label{sec:transfer}

We present off-diagonal matrix elements in
the transfer integrals of the tight-binding model.
The indices $i$ and $j$ specify a unit cell and
$l$ and $l^{\prime}$ represent a vanadium atom
in the unit cell.
Between two vanadium atoms,
the transfer matrix $t_{i j l l^{\prime} m m^{\prime}}$
has ($m, m^{\prime}$) elements, where $m$ and $m^{\prime}$
denotes an orbital.
Here we only show the transfer matrices
with at least one element larger then 0.01eV.
Other transfer matrices are also included
in the actual calculations.

\begin{itemize}
 \item $i=j$ (intra-unit cell)

       \begin{gather}
	t_{i i l l m m^{\prime}} =
	\begin{pmatrix}
	 -0.27113 & -0.00000 &  0.00000\\ 
	  0.00000 & -0.21752 &  0.00000\\
          0.00000 & -0.00000 & -0.10540
	\end{pmatrix}\\
	t_{i i 1 2 m m^{\prime}} = t_{i i 3 4 m m^{\prime}} =
	\begin{pmatrix}
	 0.1085 &  0.0407 &  0.0392\\
	 0.0407 & -0.0362 & -0.0108\\
	 0.0394 & -0.0110 &  0.1323
	\end{pmatrix}\\
	t_{i i 1 3 m m^{\prime}} = t_{i i 2 4 m m^{\prime}} =
	\begin{pmatrix}
	  0.0366 &  0.0144 &  0.0031\\
	 -0.0983 & -0.1206 & -0.0259\\
	 -0.1255 &  0.0303 & -0.0217
	\end{pmatrix}\\
	t_{i i 1 4 m m^{\prime}} = t_{i i 2 3 m m^{\prime}} =
	\begin{pmatrix}
	 0.0073 &  0.0010 & -0.0044\\
	 0.0075 &  0.0027 &  0.0067\\
	 0.0288 &  0.0047 &  0.0082
	\end{pmatrix}
       \end{gather}

 \item $\bm{r}_{i}-\bm{r}_{j} = (a, 0, 0)$
       
       \begin{gather}
	t_{i j l l m m^{\prime}} = 
	\begin{pmatrix}
	 -0.0156 & -0.0058 & -0.0077 \\
	 -0.0058 & -0.0102 &  0.0146 \\
	 -0.0078 &  0.0146 &  0.0042 
	\end{pmatrix}\\
	t_{i j 1 3 m m^{\prime}} =  t_{i j 2 4 m m^{\prime}} = t_{i i 1 3 m m^{\prime}}\\
	t_{i j 1 4 m m^{\prime}} =  t_{i j 2 3 m m^{\prime}} = 
	\begin{pmatrix}
	 -0.0145 &  0.0148 &  0.0351 \\
          0.0108 &  0.0017 &  0.0003 \\
	  0.0013 &  0.0022 & -0.0058
	\end{pmatrix} \equiv t_{i+x j 1 4 m m^{\prime}}
       \end{gather}

 \item $\bm{r}_{i}-\bm{r}_{j} = (0, b, 0)$
       
       \begin{gather}
	t_{i j l l m m^{\prime}} = 
	\begin{pmatrix}
	 -0.0052 &  0.0069 &  0.0002 \\
          0.0069 & -0.0420 & -0.0009 \\
          0.0002 & -0.0009 &  0.0202 
	\end{pmatrix}\\
	t_{i j 1 3 m m^{\prime}} =  t_{i j 2 4 m m^{\prime}} = t_{i i 1
	3 m m^{\prime}}\\
	t_{i j 1 4 m m^{\prime}} = t_{i j 2 3 m m^{\prime}} = t_{i i 1 4 m m^{\prime}}
       \end{gather}

 \item $\bm{r}_{i}-\bm{r}_{j} = (0, 0, c)$

       \begin{gather}
	t_{i j l l m m^{\prime}} = 
	\begin{pmatrix}
	 -0.0134 & -0.0003 &  0.0149 \\
	 -0.0003 &  0.0013 &  0.0040 \\
	 0.0149 &  0.0040 & -0.0092 
	\end{pmatrix}\\
	t_{i j 1 2 m m^{\prime}} = t_{i j 3 4 m m^{\prime}} = t_{i i 1 2
	m m^{\prime}}\\
	t_{i j 1 4 m m^{\prime}} = t_{i j 3 2 m m^{\prime}} = 
	t_{i+x j 1 4 m m^{\prime}}
       \end{gather}

 \item $\bm{r}_{i}-\bm{r}_{j} = (a, b, 0)$
       \begin{gather}
	t_{i j 1 3 m m^{\prime}} = t_{i j 2 4 m m^{\prime}} = 
	t_{i i 1 3 m m^{\prime}}\\
	t_{i j 1 4 m m^{\prime}} = t_{i j 2 3 m m^{\prime}}=
	t_{i+x j 1 4 m m^{\prime}}
       \end{gather}

 \item $\bm{r}_{i}-\bm{r}_{j} = (a, 0, c)$
       \begin{gather}
	t_{i j 1 4 m m^{\prime}} = t_{i i 1 4 m m^{\prime}}
       \end{gather}

 \item $\bm{r}_{i}-\bm{r}_{j} = (a, 0, -c)$
       \begin{gather}
	t_{i j 2 3 m m^{\prime}} = t_{i i 1 4 m m^{\prime}}
       \end{gather}

 \item $\bm{r}_{i}-\bm{r}_{j} = (0, b, c)$
       \begin{gather}
	t_{i j 1 4 m m^{\prime}} = t_{x+i i 1 4 m m^{\prime}}
       \end{gather}

 \item $\bm{r}_{i}-\bm{r}_{j} = (0, b, -c)$
       \begin{gather}
	t_{i j 2 3 m m^{\prime}} = t_{i+x j 1 4 m m^{\prime}}
       \end{gather}

 \item $\bm{r}_{i}-\bm{r}_{j} = (a, b,  c)$
       \begin{gather}
	t_{i j 1 4 m m^{\prime}} = t_{i i 1 4 m m^{\prime}}
       \end{gather}

 \item $\bm{r}_{i}-\bm{r}_{j} = (-a, -b,  c)$
       \begin{gather}
	t_{i j 3 2 m m^{\prime}} = t_{i i 1 4 m m^{\prime}}
       \end{gather}

\end{itemize}


\end{document}